# New Theory of Superconductivity (Part II): Confirmation at low temperature

R. Riera, J.L. Marín, R. Betancourt-Riera
Departamento de Investigación en Física, Universidad de Sonora.
Apdo. Postal 5-088, 83190
Hermosillo, Sonora, México

R. Rosas,
Departamento de Física, Universidad de Sonora.
Apdo. Postal 1626, 83000
Hermosillo, Sonora, México

An application of the general equation obtained in Part I to low critical temperature superconductors utilizing an *ad hoc* phononic theory is developed. Then, we arrive to a specific expression for the bounding energy as a function of temperature. The density of states of the electron pairs is calculated and used to obtain an equation for the critical magnetic field. This result is needed for determine the electrodynamical properties. Finally, we obtain the specific heat as a function of temperature and compare it to experimental data for $Sn$, and calculate its jump at $T_C$ for eight superconductors.

The New Theory of Superconductivity exposed in Part I is reassumed here utilizing a low temperature harmonic theory of the phonons where acoustic phonons are predominant. With the general expression obtained incorporating the parameter of superconductivity the bounding energy (energy gap) is calculated as a function of temperature. An expression for the density of states of electron pairs for this kind of superconductors is given taking into account that the energy of the local electron pair and phonon are equal. Through the use of the density of states and the energy gap we determine the relation between the critical magnetic filed and the temperature. The electrodynamic properties of those superconductors are then explained and a comparison with experimental data for the specific heat and its jump at $T_C$ is done.

For the acoustic phonons the frequency $\omega$ depends directly on the moment $q$ of such a form that $\varepsilon_p = \hbar c q$. At low temperatures $(T \ll \Theta_D)$ we first note that only phonons with $\hbar\omega(q)$ comparable to or lesser than $k_B T_C$ can be absorbed or emitted by electrons. In the case of absorption this is immediately obvious, since these are the only phonons present in appreciable number. It is also true in the case of emission that in order to emit a phonon, an electron must be far enough above the Fermi level for the final electron level (whose energy is lower by the quantity $\hbar\omega(q)$) to be unoccupied; since levels are occupied only to within order $k_B T$ above $\varepsilon_F$, and unoccupied only to within order $k_B T$ below of $\varepsilon_F$, only phonons with energies $\hbar\omega(q)$ of the order of $k_B T$ can be emitted.

Well below of Debye temperature, the condition $\hbar\omega(q) \leq k_B T_C$ requires $q$ to be small compared with $k_D$. In this regime $\omega$ is of the order of $cq$ so the wave vectors $q$ of the phonons are of the order of $k_B T_C / \hbar c$ or lesser. Thus within the surface of phonons that the conservation laws permits to be absorbed or emitted, only a subsurface of linear dimensions proportional to $T$, and hence of area proportional to $T^2$, can actually participate. We conclude that the number of phonons that can scatter an electron declines as $T^2$ becomes well below the $T_C$.

As we have already previously seen, upon arriving to the critical temperature the electron pair energy $E$ is constant and equal to the phonon energy for $q_{max}$ (see Ref. 1). Then, let make

$$E = 2\varepsilon(k) = \ell(q_{max}) = \varepsilon_p(q_{max}) = \lambda k_B T_C, \tag{1}$$

where $\lambda$ is the parameter of the superconductivity, which it is characteristic of each superconductor material and is determined from the experiments, measuring the bond energy of the electron pairs $2\varepsilon_0(T = 0K)$ and the critical temperature.

However, for $T = 0K$ we also know that $\ell(0) = 0$ since $q \approx 0$, then $E = 2\varepsilon_0(0) = \lambda k_B T_C$.

Therefore, for any temperature in the interval $0 < T < T_C$, $2\varepsilon(k) = \ell(q) = \lambda k_B T$. Taking into account the equation[1] $E^2 = \ell^2(q) + (2\varepsilon_0(T))^2$, we obtain $(2\varepsilon_0(T))^2 = (\lambda k_B T_C)^2 - (\lambda k_B T)^2$.

From here we obtain the general equation for the low temperature superconductors

$$\varepsilon_0(T) = \frac{1}{2}\lambda k_B T_C \left(1 - \frac{T^2}{T_C^2}\right)^{1/2} \quad \text{or} \quad \varepsilon_0(T) = \varepsilon_0(0)\left(1 - \frac{T^2}{T_C^2}\right)^{1/2}. \tag{2}$$

The equation (2) allow us to calculate the bond energy of the electron pairs as a function of the temperature. All the physical magnitudes of superconductors that vary with the temperature are related with this energy. Notice that for $T = T_C$, $\varepsilon_0(T_C) = 0$ and for $T = 0$, $\varepsilon_0(0) = 0$. The form suggested by Buckingham[2] is $\varepsilon_0(T) = 3.2 k_B T_C (1 - T/T_C)^{1/2}$, which it is empiric and it is not valid in $T = 0K$. The BCS[3] theory doesn't arrive to an explicit relationship for the bond energy of the electron pairs.

Some of the one-electron or lattice properties, like the specific heat and the magnetic field, are of the form

$$Q = \frac{2}{V}\sum_{n\vec{k}} Q_n(k) = 2\sum_n \int \frac{dk}{(2\pi)^3} Q_n(k) \quad \text{or} \quad Q = \frac{1}{V}\sum_{k,s} Q(\omega_s(q)) = \sum_s \int \frac{dq}{(2\pi)^3} Q(\omega_s(q)), \tag{3}$$

where for each $n$ the sum is over all the allowed $k$ given (physically distinct levels) and $s$ is the phonon branch. It is often convenient to reduce such quantities to energy or frequency integrals, introducing a density of levels or normal modes with infinitesimal energies or frequencies ranging between $\varepsilon$ and $\varepsilon + d\varepsilon$ or $\omega$ and $\omega + d\omega$

$$\int d\varepsilon g(\varepsilon) Q(\varepsilon) \quad \text{or} \quad \int d\omega g(\omega) Q(\omega). \tag{4}$$

Comparing (3) and (4) we find that density of states is given by

$$g(\varepsilon) = \sum_n \int \frac{dk}{4\pi^3} \delta(\varepsilon - \varepsilon_n(k)) \quad \text{or} \quad g(\omega) = \sum_s \int \frac{dq}{(2\pi)^3} \delta(\omega - \omega_s(q)). \tag{5}$$

In the case of electron pairs the density of states corresponding to the total energy $E$ is constant, but the corresponding to the energy of the electron pair $\ell(q)$ is variable and considering that this energy is equal to the phonon energy and it depends linearly on $q$ (notice that it does not depend on the square of $q$), then we have:

*i*) In the first case, where $E = const$ and $\ell(q_{max}) = E$, the density of states of the electron pair does not depend on $k$, then

$$g(E) = \int \frac{dk}{(2\pi)^3} \delta(E - E_{q_{max}}) = n\delta(E - E_{q_{max}}), \tag{6}$$

carrying out the integration in spherical coordinates in the reciprocate space of wave vectors, we obtain $n = k_F^3/3\pi^2 = (2/3)\varepsilon_F g(\varepsilon_F)$ with $g(\varepsilon_F) = mk_F/\hbar^2\pi^2$, but as $E$ is constant, then $\varepsilon_F = E$ and $n = (2/3)\lambda k_B T_C g(\varepsilon_F) = (2/3)Eg(\varepsilon_F)$.

*ii*) In the second case $\ell(q)$ depends linearly on $q$ and it varies from $\ell(q=0) = 0$ in $T = 0K$ until $(E^2 - \ell^2(T))^{1/2}$ for any temperature $0 < T < T_C$ or $0 < q < q_{max}$, taking into account that $\ell(q \text{ or } T) = \varepsilon_p = \hbar cq$; then, the integration of the Eq. (6) is

$$g(\ell(T)) = \frac{1}{2\pi^2} \frac{\ell^2(T)}{\hbar^3 c^3}, \tag{7}$$

where $\hbar^3 c^3 = q_F^3 \hbar^3 c^3/q_F^3 = (\varepsilon_F^3/3\pi^2)(3\pi^2/q_F^3) = \varepsilon_F^3/3n\pi^2$, and substituting the value of $n$, we obtain

$$g(\ell(T)) = \frac{1}{2\pi^2} \frac{\ell^2(T)}{\hbar^3 c^3} = \frac{\ell^2(T)}{\varepsilon_F^2} g(\varepsilon_F). \tag{8}$$

Notice that when $\ell(T) = \varepsilon_F$, then $g(\ell) = g(\varepsilon_F)$; finally we can write

$$g(\ell(T)) = \frac{\ell^2(T)}{E^2} g(\varepsilon_F), \tag{9}$$

which it is different to the obtained in the BCS[3] theory (see Eq. (3.26) of Ref. 3).

The critical magnetic field for a bulk superconductor material of unit volume it is calculated equaling the magnetic energy with the average bond energy of the electron pairs:

$$\frac{H_C^2(T)}{8\pi} = \int \varepsilon g(\varepsilon) d\varepsilon. \tag{10}$$

For $T = 0K$, $\varepsilon = E$ and $g(\varepsilon) = g(\varepsilon_F)$, the critical magnetic field can be calculated as

$$\frac{H_C^2(0)}{8\pi} = \int_0^E \varepsilon g(\varepsilon_F) d\varepsilon = g(\varepsilon_F)\frac{E^2}{2} = g(\varepsilon_F)\frac{\lambda k_B T_C}{2}, \tag{11}$$

obtaining $H_C(0) = (4\pi g(\varepsilon_F))^{1/2} \varepsilon_0(0)$. In order to calculate the critical magnetic field as a function of the temperature we use the Eq. (10) in the following form

$$\frac{H_C^2(T)}{8\pi} = \int \ell(T) g(\ell(T)) d\ell; \tag{12}$$

substituting the expression for the density of states given by Eq. (9), we obtain

$$\frac{H_C^2(T)}{8\pi} = \int \ell(T) \frac{\ell^2(T)}{E^2} g(\varepsilon_F) d\ell = 2g(\varepsilon_F) \frac{\ell^4(T)}{4E^2}\bigg|_0^{(E^2-\ell^2)^{1/2}}, \tag{13}$$

and this way we have obtained that

$$\frac{H_C^2(T)}{8\pi} = = 2g(\varepsilon_F)\frac{\varepsilon_0^4(T)}{4E^2}. \tag{14}$$

Substituting the Eq. (2) in Eq. (14), we obtain

$$H_C(T) = \sqrt{4\pi g(\varepsilon_F)}(\lambda k_B T_C)\left[1 - \frac{T^2}{T_C^2}\right] = H_C(0)\left(1 - \frac{T^2}{T_C^2}\right). \tag{15}$$

This expression is valid in the whole interval of temperature $0 \leq T \leq T_C$ and it coincides with the empirical law obtained from the experiments; in this case the BCS theory deduces an expression for the critical magnetic field only in $T = 0K$.

The electrodynamical properties of the superconductors are closely related or very related with the critical magnetic field considered in the Eq. (15). Taking into account the Maxwell equation

$\nabla \times \boldsymbol{H} = (4\pi/c)\boldsymbol{J}$ and applying the rotor in both members and considering that $\nabla \cdot \boldsymbol{H} = 0$, we obtain $\nabla^2 \boldsymbol{H} = -(4\pi/c)\nabla \times \boldsymbol{J}$. Now if we consider that an electric force of the form $d\boldsymbol{p}/dt = e\boldsymbol{E}$ is generated, and considering that $\boldsymbol{J} = ne\boldsymbol{v}$, we obtain $(m/ne^2)(d\boldsymbol{J}/dt) = \boldsymbol{E}$; then, applying the rotor in both members and considering the Maxwell equation $\nabla \times \boldsymbol{E} = -(1/c)(d\boldsymbol{H}/dt)$, we obtain that $(cm/ne^2)\nabla \times \boldsymbol{J} = -\boldsymbol{H}$. Now if we make $\Lambda = 4\pi\lambda_L^2/c^2 = (m/ne^2)$ and substituting in the previous equation we obtain one the London equations[4] $\boldsymbol{H} = -c\nabla \times \Lambda\boldsymbol{J}$, which when it is combined with $\nabla^2 \boldsymbol{H} = -(4\pi/c)\nabla \times \boldsymbol{J}$ leads to the second London equation $\nabla^2 \boldsymbol{H} = \boldsymbol{H}/\lambda_L^2$. This implies that a magnetic field is exponentially screened from the interior of a sample and only can penetrate the length $\lambda_L$, this is the Meissner effect. Thus, the parameter $\lambda_L$ is operationally defined as a penetration depth.

From the equation $(cm/ne^2)\nabla \times \boldsymbol{J} = -\boldsymbol{H}$ and by using the relation $\boldsymbol{H} = \nabla \times \boldsymbol{A}$ we obtain $\boldsymbol{J} = -(ne^2/mc)\boldsymbol{A} = -(1/\Lambda c)\boldsymbol{A}$, this is the diamagnetic density of current, which is valid for $T = 0K$; therefore we can write it of the following form

$$\boldsymbol{J}(0) = -\frac{1}{c\Lambda(0)}\boldsymbol{A}(0) = \boldsymbol{J}_D(0). \tag{16}$$

In order to calculate the paramagnetic density of current we use the equation (15) and substituting the equation $\boldsymbol{H} = -c\nabla \times \Lambda\boldsymbol{J}$ we obtain

$$-c\Lambda\nabla \times \boldsymbol{J}(T) = \boldsymbol{H}_C(0)\left(1 - \frac{T^2}{T_C^2}\right) = -c\Lambda(0)\left(1 - \frac{T^2}{T_C^2}\right)\nabla \times \boldsymbol{A}(0), \tag{17}$$

where

$$\Lambda(T) = \Lambda(0)\left(1 - \frac{T^2}{T_C^2}\right). \tag{18}$$

This equation coincides with the equation (5.25) of the paper of BCS theory[3]

$$\Lambda(T) = \Lambda(0)\left(1 + \frac{\beta}{\varepsilon_0(T)} \frac{d\varepsilon_0(T)}{d\beta}\right), \text{ with } \beta = \frac{1}{k_B T},$$ if we use the equation (2) for the electron pairs bond energy $\varepsilon_0(T)$.

The paramagnetic current is

$$\boldsymbol{J}_p(T) = \frac{1}{c\Lambda(T)} \boldsymbol{A}(0). \tag{19}$$

Notice that the paramagnetic density of current contains the diamagnetic density of current, from such a form that the total induced density of current is

$$\boldsymbol{J} = \boldsymbol{J}_D + \boldsymbol{J}_p = -\frac{1}{c\Lambda(0)} \boldsymbol{A} + \frac{1}{c\Lambda(T)} \boldsymbol{A} = -\frac{1}{c\Lambda(T)} \boldsymbol{A}, \tag{20}$$

then, using the Eq. (19) we obtain the Eq. (5.26) of the paper of BCS[3] theory:

$$\boldsymbol{J} = \boldsymbol{J}_D + \boldsymbol{J}_p = -\frac{\Lambda(0)}{\Lambda(T)} \frac{ne^2}{mc} \boldsymbol{A}(0).$$

Now, considering the equation $\nabla^2 \boldsymbol{H} = \boldsymbol{H}/\lambda_L^2$ and substituting $\boldsymbol{H}_C(T)$, we obtain

$$\nabla^2 \boldsymbol{H}_C(T) = \frac{1}{\lambda_L^2(0)}\left(1 - \frac{T^2}{T_C^2}\right)\boldsymbol{H}_C(0); \tag{21}$$

then making

$$\frac{1}{\lambda_L^2(T)} = \frac{1}{\lambda_L^2(0)}\left(1 - \frac{T^2}{T_C^2}\right) \tag{22}$$

we obtain that $\lambda_L(T)$ is given by the following expression

$$\lambda_L(T) = \lambda_L(0)\left(1 - \frac{T^2}{T_C^2}\right)^{-1/2}, \tag{23}$$

which it is in correspondence with our equation for $\Lambda(T)$ and with $\lambda_L(T)$ from the BCS theory; however it does not coincide with the empirical law

$$\lambda_L(T) = \lambda_L(0)\left(1 - \frac{T^4}{T_C^4}\right)^{-1/2}. \tag{24}$$

Taking the relationship between the phonon moment $q$ and the coherence distance $\xi_0$

$$\xi_0 = \frac{1}{\pi q} \tag{25}$$

and knowing that in $T = 0K$, $2\varepsilon_0 = \lambda k_B T_C = \hbar c q_{max}$, then $q = \frac{\lambda k_B T_C}{\hbar c}$, and after substituting in the Eq. (25) we obtain

$$\xi_0(0) = \frac{1}{\pi}\frac{\hbar c}{\varepsilon_0(0)} = \frac{1}{\pi}\frac{1}{\lambda}\frac{\hbar c}{k_B T_C}. \tag{26}$$

In the BCS theory $1/\lambda\pi = 0.18$, but Faber and Pippard[5] obtained a value of 0.15 and Glove and Tinkhan[6] obtained a value of 0.27. In order to calculate the dependence on the temperature of $\xi_0(T)$ we take into account the equation $2\varepsilon_0(T) = \hbar c q(T)$, where

$$q(T) = \frac{\lambda k_B T_C}{\hbar c}\left(1 - \frac{T^2}{T_C^2}\right)^{-1/2}, \tag{27}$$

obtaining finally that

$$\xi_0(T) = \frac{\hbar c}{\pi\lambda k_B T_C}\left(1 - \frac{T^2}{T_C^2}\right)^{-1/2}. \tag{28}$$

Taking into account the Eq. (24) for $\lambda_L(T)$ we obtain that

$$\kappa = \frac{\lambda_L(T)}{\xi_0(T)} = \frac{\hbar c \lambda_L(0)}{\pi\varepsilon_0(0)}, \tag{29}$$

the ratio of the two characteristic lengths defined in the Ginzburg-Landau[7] theory, which it is independent of the temperature, but it is characteristic for each superconductor.

Upon arriving to $T_C$, we can continue with our starting points in the general theory of the superconductivity, where the electron pairs energy is equal to the phonon energy. We can say that physically the specific heat of the free electrons becomes in the specific heat of the electron pairs and then in the specific heat of the phonons, therefore we should make a theory of the specific heat very similar to the phonons one, but using the Maxwell-Boltzmann distribution function and introducing the superconductivity parameter. The low temperature specific heat of the phonons in the quantum theory of the harmonic solid is given by

$$c_v = \frac{\partial}{\partial T} \sum_s \int \frac{d\mathbf{q}}{(2\pi)^3} \frac{\hbar\omega_s(\mathbf{q})}{e^{\frac{\hbar\omega_s(\mathbf{q})}{k_B T}} - 1} . \tag{30}$$

As we can see, at very low temperature modes with $\hbar\omega_s(\mathbf{q}) \gg k_B T$ will contribute negligibly to the specific heat, since the integrand will vanish exponentially. However, because $\omega_s(\mathbf{q}) \to 0$ as $\mathbf{q} \to 0$ in the three acoustic branches, these conditions will fail to be satisfied by acoustic modes of sufficiently long-wavelength, no matter how low is the temperature. These modes (and only these) will continue contributing appreciably to the specific heat.

Bearing this in mind, we can make the following simplifications in Eq. (30), all of which result in a vanishing small fractional error, in the zero-temperature limit: i) Even if the crystal has a polyatomic basis, we can ignore the optical modes in the sum over $s$, since their frequencies are bounded below; ii) We can replace the dispersion relation $\omega = \omega_s(\mathbf{q})$ for the three acoustic branches $k_B T/\hbar$ because they are substantially lesser than those frequencies at which the acoustic dispersion curves begin to differ appreciably from their long-wavelength linear forms; iii) We will substitute the Bose-Einstein distribution function for the Maxwell-Boltzmann distribution function; thus at low temperature for the superconductor electron pairs the Eq. (30) may be simplified to

$$c_v = \frac{\partial}{\partial T} \sum_s \int \frac{d\mathbf{q}}{(2\pi)^3} \hbar\omega_s(\mathbf{q}) e^{-\frac{\hbar\omega_s(\mathbf{q})}{k_B T}} , \tag{31}$$

where the integral is over the $q$ values in the interval $0 < q < q_{max}$, which it corresponds to the temperature interval $0 < T < T_C$. We evaluate the integral in spherical coordinates, writing $d\mathbf{q} = q^2 dq d\Omega$ and making the sum on $s$, we obtain

$$c_v = \frac{3}{2\pi^2} \frac{\partial}{\partial T} \int \hbar c q^3 e^{-\frac{\hbar c q}{k_B T}} dq. \tag{32}$$

If we make the change of variables $\ell = \hbar c q$ and considering that $\partial/\partial T = (\partial \ell/\partial T)(\partial/\partial \ell)$ and $\ell = \lambda k_B T$ we obtain

$$c_v = \frac{3}{2\pi^2} \frac{1}{\hbar^3 c^3} \lambda k_B T \frac{d}{d\ell} \int_0^{\ell(T)} \ell^3 e^{-\frac{\ell}{k_B T}} d\ell; \tag{33}$$

previously we have obtained that $\hbar^3 c^3 = E^2/2\pi^2 g(\varepsilon_F)$ and considering that $E = \lambda k_B T_C$ and $\ell = \lambda k_B T$, we obtain

$$c_v = 3g(\varepsilon_F) k_B^2 \lambda^2 \left(\frac{T}{T_C}\right)^3 (1 - e^{-\lambda}), \tag{34}$$

and as the electronic specific heat in $T \geq T_C$ is

$$c_v^e(T = T_C) = \frac{2\pi^2}{3} k_B^2 g(\varepsilon_F) T_C. \tag{35}$$

Taking into account that $\lambda = 2\varepsilon_0(0)/k_B T_C$ we finally obtain for the specific heat in the superconducting state the following expression:

$$\frac{c_v}{c_v^e(T_C)} = \frac{2}{\pi^2} \left(\frac{T}{T_C}\right)^3 \left(\frac{2\varepsilon_0(0)}{k_B T_C}\right)^2 \left(1 - e^{-\frac{2\varepsilon_0(0)}{k_B T_C}}\right). \tag{36}$$

This expression is valid in the whole interval of temperature $0 < T \leq T_C$ and it describes the jump of the specific heat in $T = T_C$, taking into account the particular characteristic of each superconductor material through the parameter $\lambda$ of the superconductivity. This expression can be written as $c_v/c_v^e(T_C) = \zeta(T/T_C)^3$, where

$$\zeta = (2/\pi^2)\lambda^2(1-e^{-\lambda}), \tag{37}$$

which it can be called a *characteristic parameter of the specific heat*.

The ratio $c_v/c_v^e$ vs $T/T_C$ is plotted in Fig. 1 (using Eq. (36)) and it is compared with experimental values for tin; we used $\lambda = 3.5 \pm 0.1$ reported in Ref. 8. The agreement of our expression is rather good. In Table 1 we show measured values of the ratio $(c_v - c_v^e)/c_v^e\big|_{T_C}$ reported in Ref. 8 and calculated for our equation.

We think the problems that have presented the various theories of superconductivity are due to an incorrect physical interpretation of this phenomenon and of the energies appearing in the general equation of superconductivity. Besides, those theories don't keep in mind that the electron pair energies are equal to the phonon energies.

The superconductivity can only occur in materials where free electric charges exist, such like the conductors. The superconductivity in pure dielectrics cannot exist, because in those materials the electric charges don't exist, so that, for superconductivity occurs should be occurs a conductor-superconductor transition of phase. Due to this condition, if we want to obtain a superconductor with high critical temperatures, then we need an electron-phonon coupling constant strong enough to have electron pairs correlated with small coherence distance; this is to say very near electrons. In order to obtain strong coupling between the electrons and the phonons we need materials like the dielectrics, but we know that in those materials there are not free electric carriers, therefore, we need to introduce conductor or metallic phases inside them with the purpose of free carriers can move through that phases as it happens in the high $T_c$ ceramic superconductors. In conclusion the ideal superconductor should have a big electron-phonon coupling constant, which it is characteristic of the dielectrics and conductor phases where the free charges exist and they can form the electron pairs.

The relation $2\varepsilon_0(0) = 3.5 k_B T_C$ imposed in the BCS theory[3] forbids the existence of superconductivity in high $T_C$. We think this relation is important and it should be rewritten in the following form: $2\varepsilon_0 = \lambda k_B T_C$, where $\lambda$ is our superconducting parameter, which it depends on each type of superconductor and it can be experimentally measured. For the superconductors characterized by high $T_C$ this parameter is big.

It is physically acceptable that an electron pair is a new quasi-particle, which it is formed by two electrons and one phonon occupying the same states, but without losing any of their own identity, the electrons continue being fermions and the phonon a boson; that is the reason because we use the statistics of Maxwell-Boltzmann.

If the hypothesis of the formation of electron pairs is experimentally proven in any type of superconductivity, then the electron-electron Coulomb repulsion should compensate any interaction responsible for the electric resistance, which in this case it is considered as a different type of interaction from the electron-phonon one.

Caption of figures

Fig. 1. Ratio $c_v/c_v^e$ vs $T/T_C$

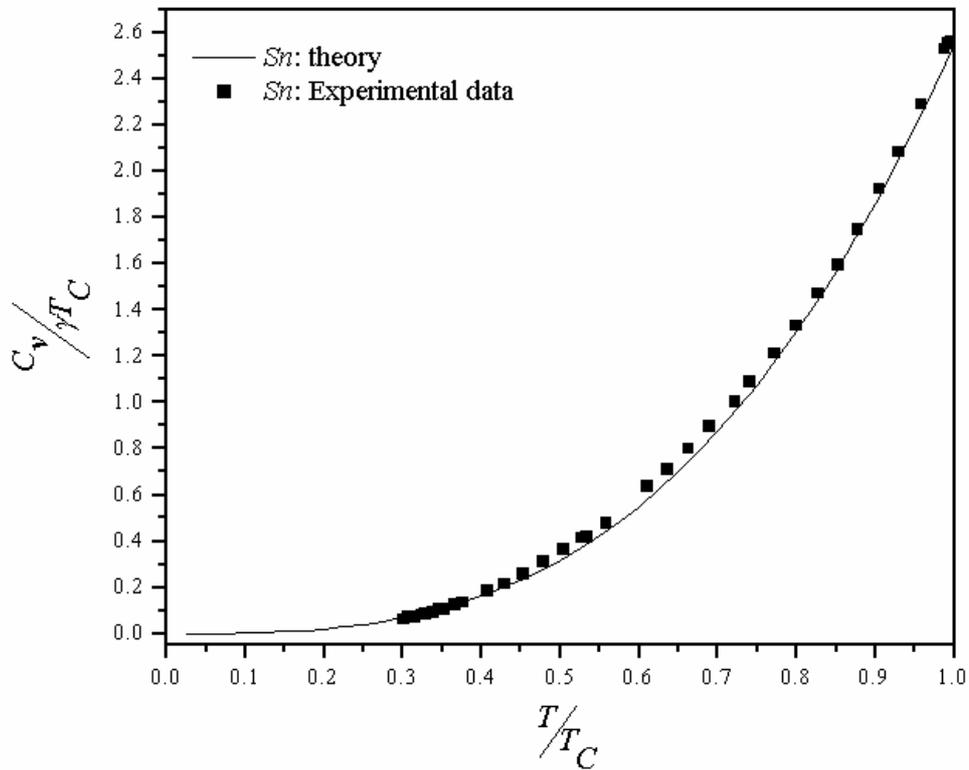

Fig.1

Table 1. We show measured values of the ratio $(c_v - c_v^e)/c_v^e\big|_{T_C}$ reported in Ref. 8 and calculated for our equation. We have considered the values listed of the λ parameter of Ref. 8 which they have an uncertainly ±.1. The values in bold and underlined are in good agreement with the experiment data, considering the uncertainty in λ

| ELEMENTS | SUPERCONDUCTIVITY PARAMETERS $\lambda = \dfrac{2\varepsilon_0(0)}{k_B T_C} \pm 0.1$ | EXPERIMENT DATA $(c_v - c_v^e)/c_v^e\big|_{T_C}$ | THEORETICAL DATA $(c_v - c_v^e)/c_v^e\big|_{T_C}$ | | |
|---|---|---|---|---|---|
| | | | -0.1 | λ | +0.1 |
| Al | 3.4 | 1.4 | 1.12 | 1.26 | **<u>1.41</u>** |
| Nb | 3.8 | 1.9 | | **<u>1.86</u>** | |
| Pb | 4.3 | 2.7 | | **<u>2.69</u>** | |
| Sn | 3.5 | 1.6 | 1.26 | 1.41 | **<u>1.55</u>** |
| V | 3.4 | 1.5 | 1.12 | 1.26 | **<u>1.41</u>** |
| Ta | 3.6 | 1.6 | 1.41 | **<u>1.55</u>** | 1.70 |
| Tl | 3.6 | 1.5 | 1.41 | **<u>1.55</u>** | 1.70 |
| In | 3.6 | 1.7 | 1.41 | 1.55 | **<u>1.70</u>** |